\documentclass[a4paper,11pt]{article}

\usepackage{jinstpub} 
\include{00README.XXX}

\title{Status of installation and commissioning for the Belle II time-of-propagation counter}

\author[a]{Y. Maeda,\note{Corresponding author.}}

\affiliation[a]{Kobayashi-Maskawa Institute, Nagoya University,\\Furo, Chikusa, Nagoya, Aichi, Japan}

\emailAdd{maeday@hepl.phys.nagoya-u.ac.jp}

\abstract{
The Time-Of-Propagation (TOP) counter is a novel device for particle identification
for the barrel region of the Belle II experiment,
where, information of Cherenkov light propagation time is used to reconstruct its ring image.
We successfully finished the detector production and installation to the Belle II structure in 2016.
Commissioning of the installed detector has been on going,
where the detector operation in the 1.5-T magnetic field was studied.
Although we found a problem where photomultipliers were mechanically moved due to the magnetic force,
it was immediately fixed.
Performance was evaluated with cosmic ray data,
the number of photon hits was confirmed to be consistent with simulation within 15$-$30\%. 
}


\arxivnumber{1705.09895} 

\collaboration[c]{on behalf of the Belle II TOP group}

\proceeding{Instrumentation of Colliding Beam Physics (INSTR-17) \\
  27 February 2017 - 3 Mar 2017\\
  Budker Institute of Nuclear Physics, Novosibirsk, Russia}

\begin{document}
\maketitle
\flushbottom

\section{Introdution}
\label{sec:intro}

\begin{figure}[tb]
\begin{minipage}{0.48\textwidth}
\includegraphics[width=0.8\textwidth,bb=0 0 330 241,clip]{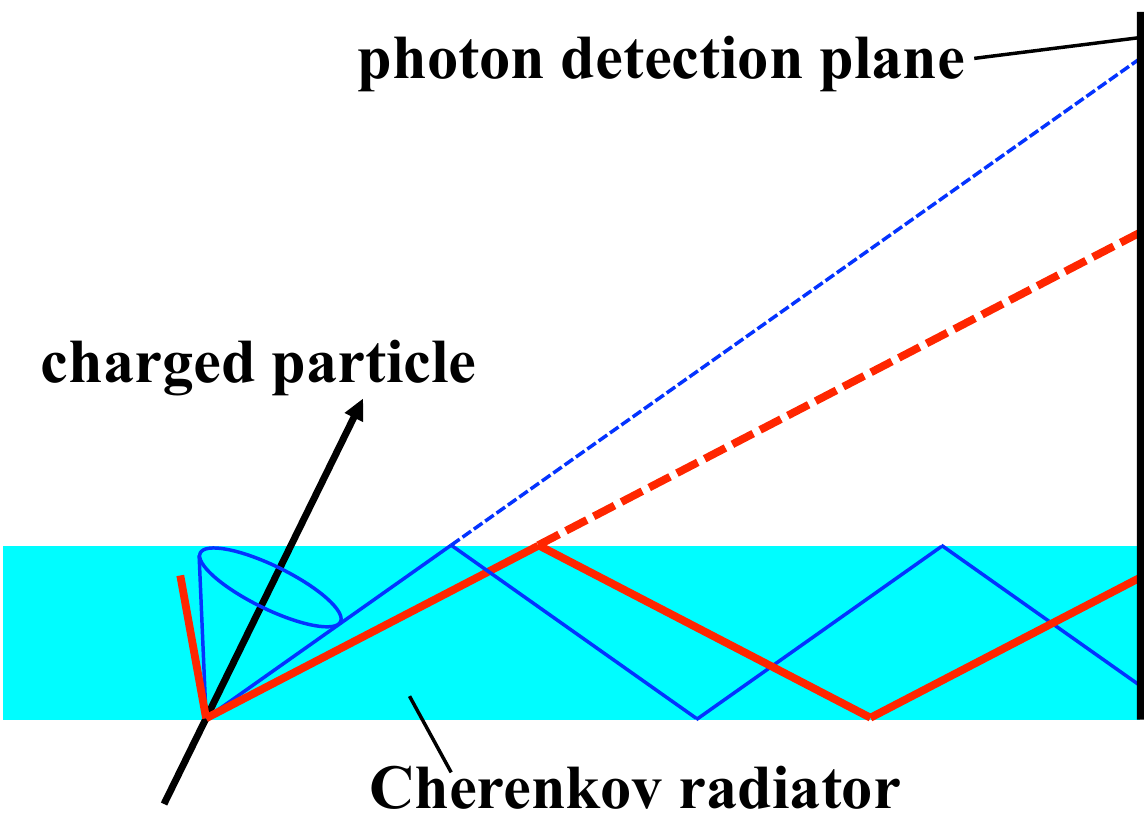}
\caption{\label{fig:Principle} Principle of particle identification in the TOP counter.}
\end{minipage}
\begin{minipage}{0.04\textwidth}
\end{minipage}
\begin{minipage}{0.48\textwidth}
\includegraphics[width=0.87\textwidth,bb=0 0 410 276,clip]{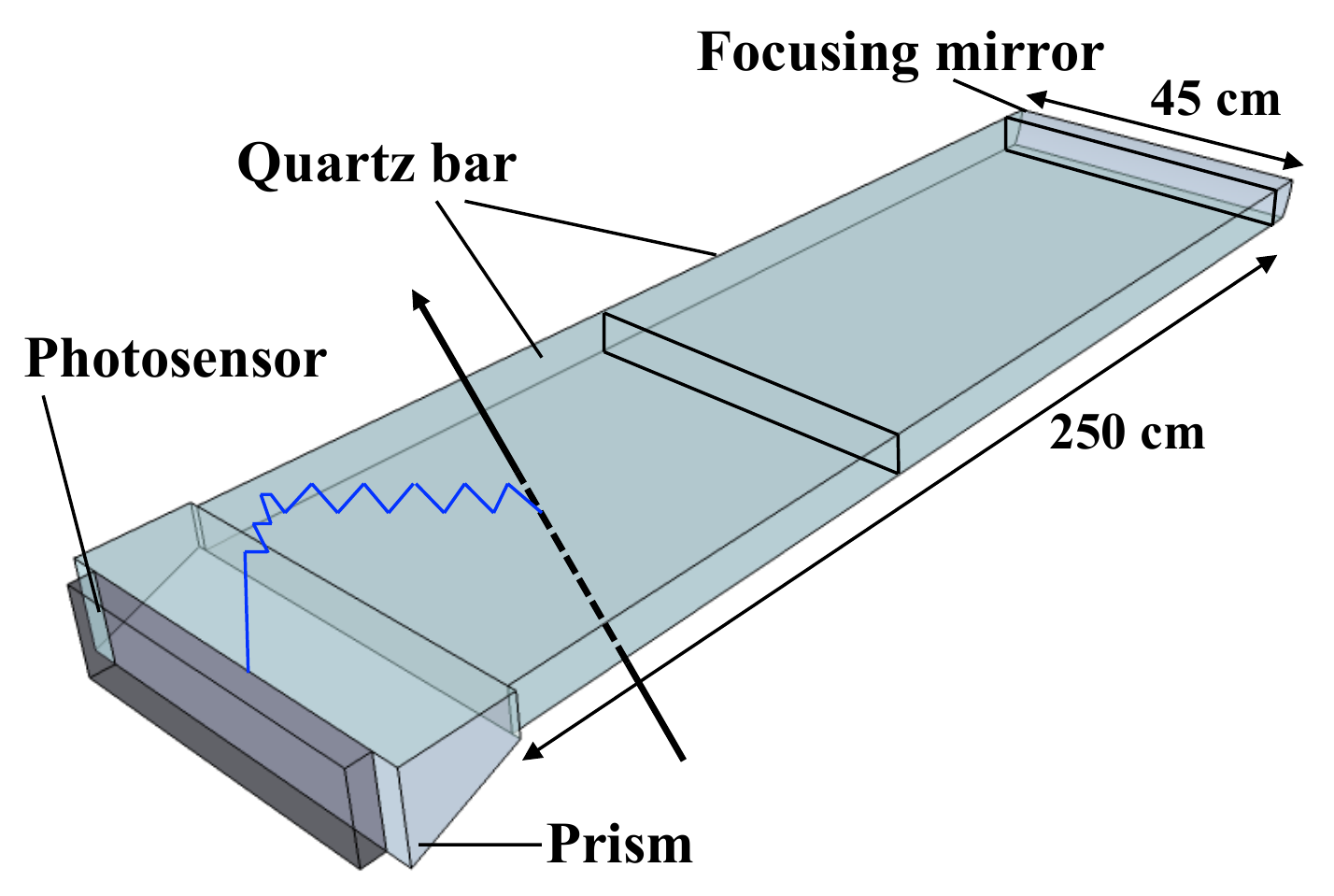}
\caption{\label{fig:Optics} Overview of optical system of the TOP counter.}
\end{minipage}
\end{figure}

Upgrade of the particle identification (PID) system from the Belle experiment
is one of the key points to achieve better precision in a search for new physics in the Belle II experiment. 
The Time-Of-Propagation (TOP) counter is one of the two new detectors dedicated for PID,
which covers the barrel region of the Belle II detector.
The detector is based on a novel idea to use timing information to reconstruct a ring image of Cherenkov light.
A particle can be identified by measuring propagation time of Cherenkov light to bar ends,
because its emission angle depends on the particle velocity
and it gives difference in path length as shown in Fig.~\ref{fig:Principle}. 
Compared to the threshold-type PID detectors used in the previous Belle experiment \cite{BelleACC},
significant improvement of PID performance is expected thanks to ring image information.
This concept also allows the detector to be more compact, light and homogeneous,
which helps to improve performance of the tracking detector (Central Drift Chamber, CDC) \cite{CDC}
and electromagnetic calorimeter (ECL) \cite{ECL},
thanks to larger available volume and less scattering of low momentum particles, respectively.

\section{Detector design}
\subsection{Optical system}

The TOP counter consists of 16 identical modules,
which forms a barrel structure to cover the central region of the Belle II detector.
Each module consists of optical system, photosensors, readout electronics
and a support structure to hold these components.
The optical system is a combination of quartz bars as Cherenkov radiator and propagator,
a focusing mirror and a prism. 
Each quartz bar has dimensions of 125~cm in length, 45~cm in width and 2~cm in thickness,
and two bars are glued to make a single 2.5-m-long bar.
The refractive index is 1.47 at wavelength of 400~nm.
Those quartz bars need to meet various challenging requirements,
such as surface roughness ($<$5~\r{A}), parallelism ($<$4~arcsec) and flatness ($<$6.3~$\mu$m).
Quality of each bar was assessed before assembly,
and we confirmed all the bars satisfied the requirements
of bulk transmittance ($>98.5\%/$m) and internal reflectance ($>99.9$\%).
The glued quartz bar is stored in a support structure made of aluminum honeycomb plates,
which owns enough rigidity with light material \cite{mechanics}.

\subsection{Photosensor}

Cherenkov light propagating inside the quartz bar is finally detected by photosensors, 
which are attached to the prism surface.
We have successfully developed and produced
more than 500 square-shaped micro-channel-plate photo multipliers (MCP-PMTs) for this detector \cite{MCPPMT}.
In this PMT, a single photon is detected with timing resolution better than 50~ps
that allows us to distinguish pions and kaons of multi GeV/$c$ momentum,
where the difference of photon propagation time is as small as an order of 100~ps.
Its anode is divided into 4$\times$4 channels.
For each TOP module, 32 PMTs are arrayed in a 2$\times$16 grid,
and attached to the prism surface via transparent silicon rubber.
The fraction of sensitive area is 73\%, thanks to the square shape of the PMTs.

Lifetime of MCP-PMT, or degradation of quantum efficiency, is an issue in using this type of PMT
under an environment of high radiation from the accelerator.
Various modifications have been applied to reduce effect of neutral gas and ion feedback,
which is considered to damage the photocathode and deteriorate quantum efficiency
according to accumulated output charge.
Several types of PMTs have been developed to satisfy the requirements.
Lifetime of the latest type is longer than 13.6~C/cm$^{2}$,
which corresponds to longer lifetime than the requirement by a factor of 3.7 \cite{MCPPMT}.

\section{Status of detector production, installation, and commissioning}
\subsection{Production and installation}

The production of the real detector modules was started in late 2014
and 17 modules, including one spare, were produced by April, 2016.
The produced modules were tested one by one with a laser calibration system \cite{Laser}
and cosmic ray data before installation.

Each tested module was installed using movable stages,
where a guide pipe was supported by the stages and a module was held along the guide pipe
so that it was able to move in any directions as well as rotate around the guide pipe.
During the installation process, module deflection was monitored using 3 types of deflection sensors.
Module deflection during the process was smaller than the requirement of 0.5~mm.
Installation of all the modules completed in May, 2016 as shown in Fig.\ref{fig:InstallationCompleted}. 
Detailed procedures are described in Ref.~\cite{mechanics}.
 
\subsection{Timing calibration}

\begin{figure}[tb]
\begin{minipage}{0.48\textwidth}
\centering
\includegraphics[width=0.9\textwidth,bb=0 0 385 290,clip]{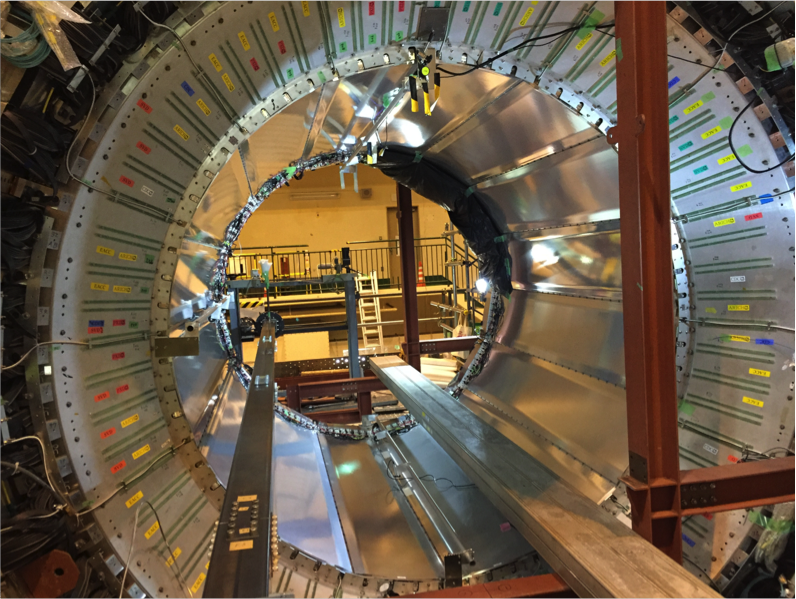}
\caption{\label{fig:InstallationCompleted} Photograph after the installation of all the 16 modules.}
\end{minipage}
\begin{minipage}{0.04\textwidth}
\end{minipage}
\begin{minipage}{0.48\textwidth}
\includegraphics[height=0.6\textwidth,bb=10 5 520 350,clip]{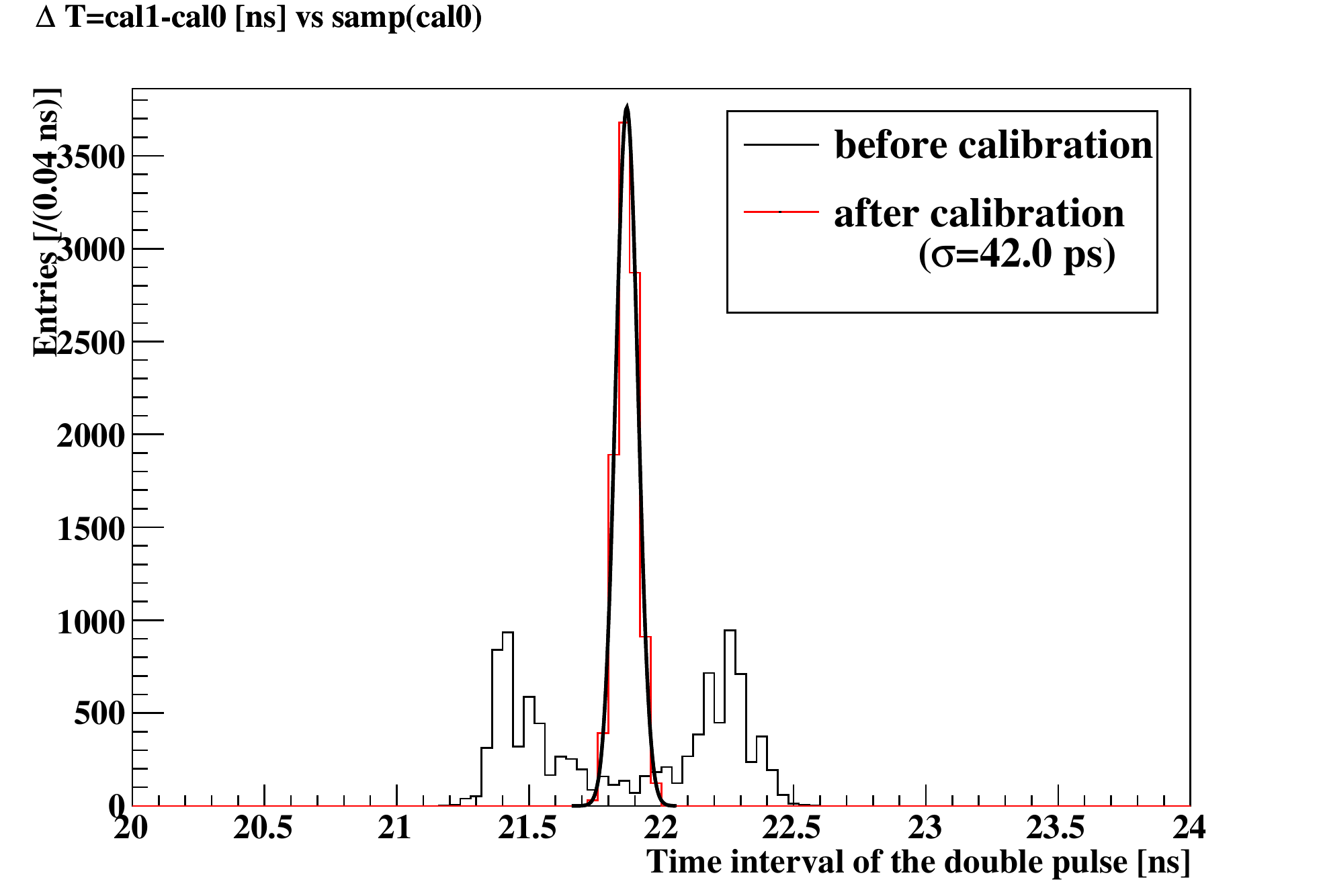}
\caption{\label{fig:TBCResult} An example of sampling time calibration result.
		Before calibration, uniform sampling intervals are assumed.}
\end{minipage}
\end{figure}

Signals from the MCP-PMTs were digitized through custom-made electronics
and waveform data with a sampling rate of 2.7~GHz was recorded for each channel.
The sampling intervals are not completely uniform,
and they must be calibrated to obtain required timing resolution.
This was achieved using test signals consisting of double pulses with a constant interval,
where the test signal was injected at random timing for each sample number
and each sampling interval was tuned
so that the double pulse interval was always constant \cite{calib}.
An example of this calibration is shown in Fig.~\ref{fig:TBCResult},
where resolution of the double pulse interval was obtained as $\sigma=42$~ps.
Contribution to timing resolution from readout electronics is estimated
to be $\sigma/\sqrt{2}\sim30$~ps.
After calibrating each sampling interval,
timing resolution for laser photon signals was examined.
With simple analysis method of hit timing calculation, $\sigma=120$~ps is obtained.
Development of algorithm for better timing resolution, which involves waveform fitting, is in progress.

\subsection{Operation in the magnetic field}

The Belle II solenoid was turned on in June and July, 2016,
where the whole TOP system was operated in the same magnetic field of 1.5 T with the real experiment for the first time. 
In this test, signals from several PMTs were found to be lost.
This was immediately turned out to be derived from the magnetic force to the housings of the PMTs.
These PMTs rotated by the force,
which resulted in loss of electrical contact to the frontend electronics
and optical coupling to the prism surface.
This ``PMT rotation'' problem was solved by inserting a plastic plate with thickness of $1.3-1.5$~mm
between a PMT holder and TOP module structure to prevent the PMT from moving.

\subsection{Performance study with cosmic ray} \label{sec:CosmicRayTest}

\begin{figure}[tb]
\begin{minipage}{0.48\textwidth}
\vspace{-1.6cm}
\includegraphics[width=0.8\textwidth,bb=0 0 703 754,clip]{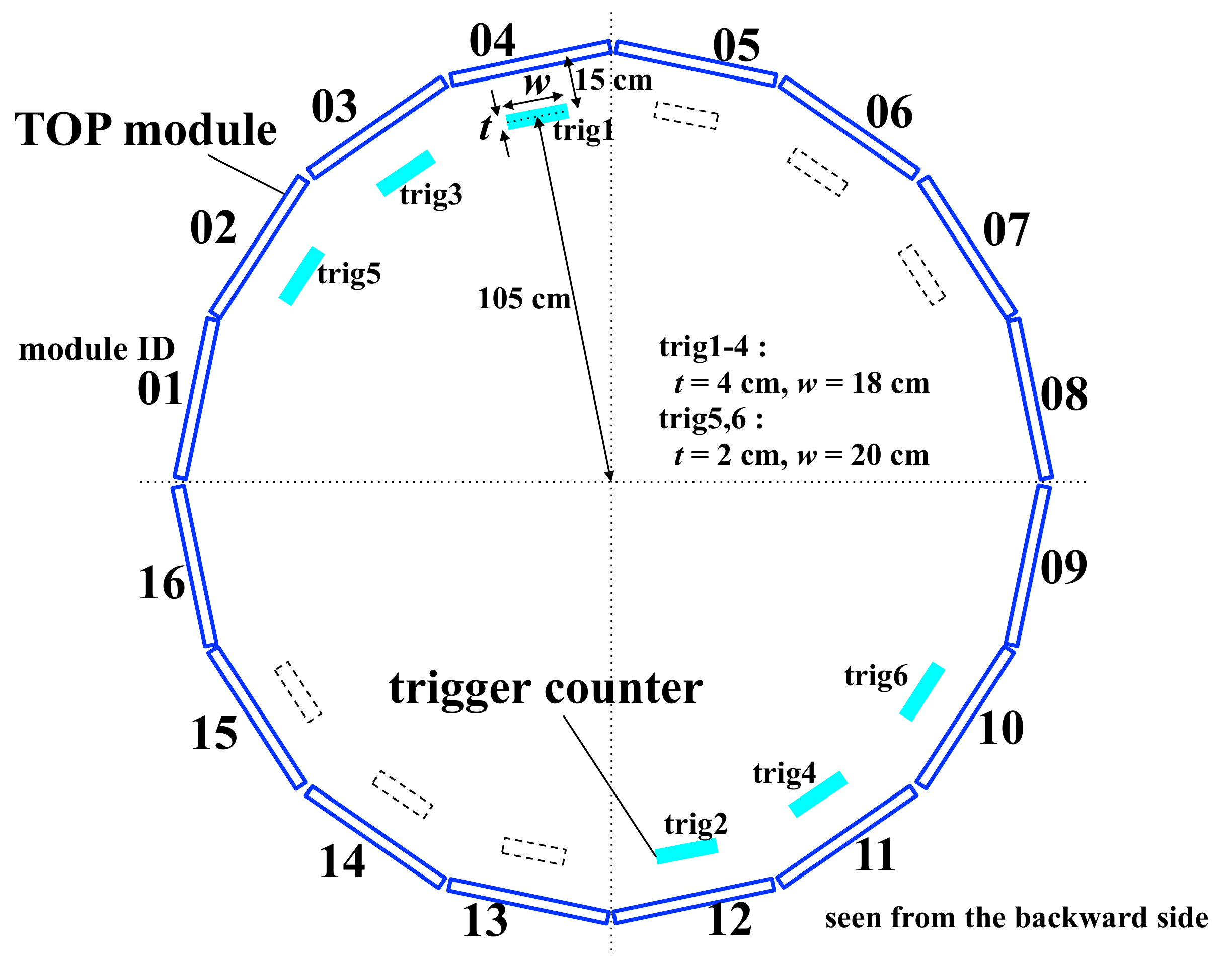}
\caption{\label{fig:CosmicTrigLayout} Layout of the TOP modules and the cosmic trigger counters.
		Configuration in taking data for module ID 02-04 and 10-12 is shown.
		The boxes with dashed line indicate the trigger counter position for other modules (05-07 and 13-15).}
\end{minipage}
\begin{minipage}{0.04\textwidth}
\end{minipage}
\begin{minipage}{0.48\textwidth}
\vspace{0.48cm}
\includegraphics[width=0.95\textwidth,bb=0 0 359 253,clip]{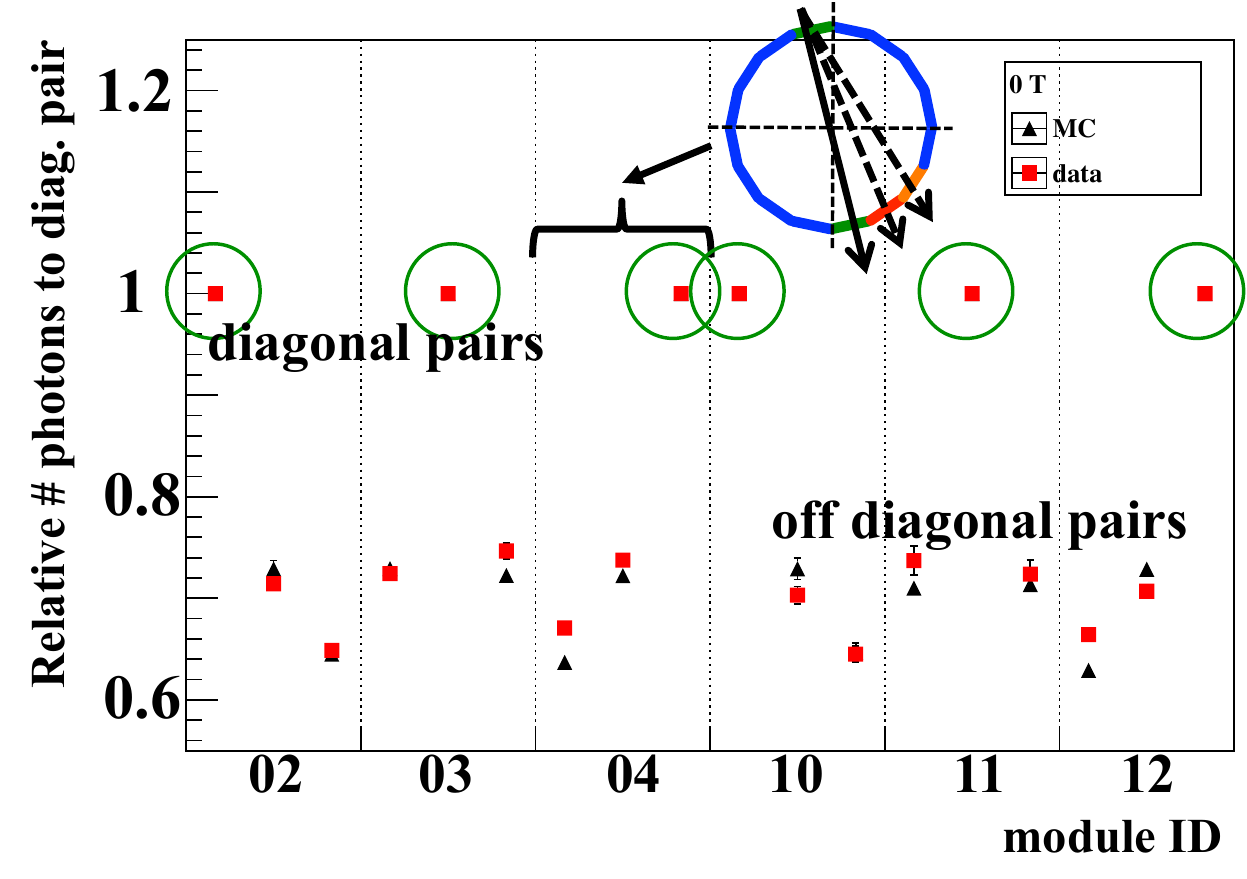}
\caption{\label{fig:CosmicOffDiag}
		Data-MC comparison of relative $N_{\mathrm{hit}}$ of events with hits in off-diagonal pairs of modules
		to diagonal-pair modules.
		The result is based on data taken for zero magnetic field.}
\end{minipage}
\end{figure}

After fixing the PMT rotation problem, we took cosmic ray data to validate detector performance
in the magnetic field.
Each TOP module was equipped with a cosmic ray trigger counter as shown in Fig.~\ref{fig:CosmicTrigLayout},
which consisted of a plastic scintillator bar with dimensions of 40~cm in length and $18-20$~cm in width.
In total 6 counters were prepared.
Their position along the beam axis was at the same with the collision point. 
Fine-mesh PMTs from the Belle TOF detector \cite{BelleTOF} were used in the trigger counters 
so that they can be operated in the magnetic field.

For performance evaluation, as track information was not available,
the number of observed photon hits ($N_{\mathrm{hit}}$) in each TOP module
was compared between data and Monte Carlo (MC) simulation,
where generation and tracing of Cherenkov photons were managed by the Geant4 package \cite{geant4}.
Cosmic ray events were triggered
based on coincidence of hits in one counter from the upper side and one counter from the lower side.
In the simulation, initial cosmic ray momentum and angle distribution was generated
based on an equation described in Ref.~\cite{CosmicFluxFormula} with measurement results given by Ref.~\cite{BESS_flux}.
Events with a single track were then selected by requiring $N_{\mathrm{hit}}>10$ for the slot
which is located in the diagonal or close-to-diagonal position to the slot in interest
and no hits ($N_{\mathrm{hit}}<10$) in all the other slots.

Figure~\ref{fig:CosmicOffDiag} shows relative $N_{\mathrm{hit}}$ values
for several impact angles to the quartz bar,
normalized to the perpendicularly incident case with hits in the diagonal module.
With a shift of particle impact angle from 90$^{\circ}$,
a part of Cherenkov photons can escape from the quartz bar
because they have larger angle than the critical angle in reflection.
This resulted in lower $N_{\mathrm{hit}}$ values when we required hits in off-diagonal modules. 
The loss of photons is well reproduced in the MC simulation.

\begin{figure}[tb]
\begin{minipage}{0.49\textwidth}
\includegraphics[width=0.95\textwidth,bb=0 0 364 249,clip]{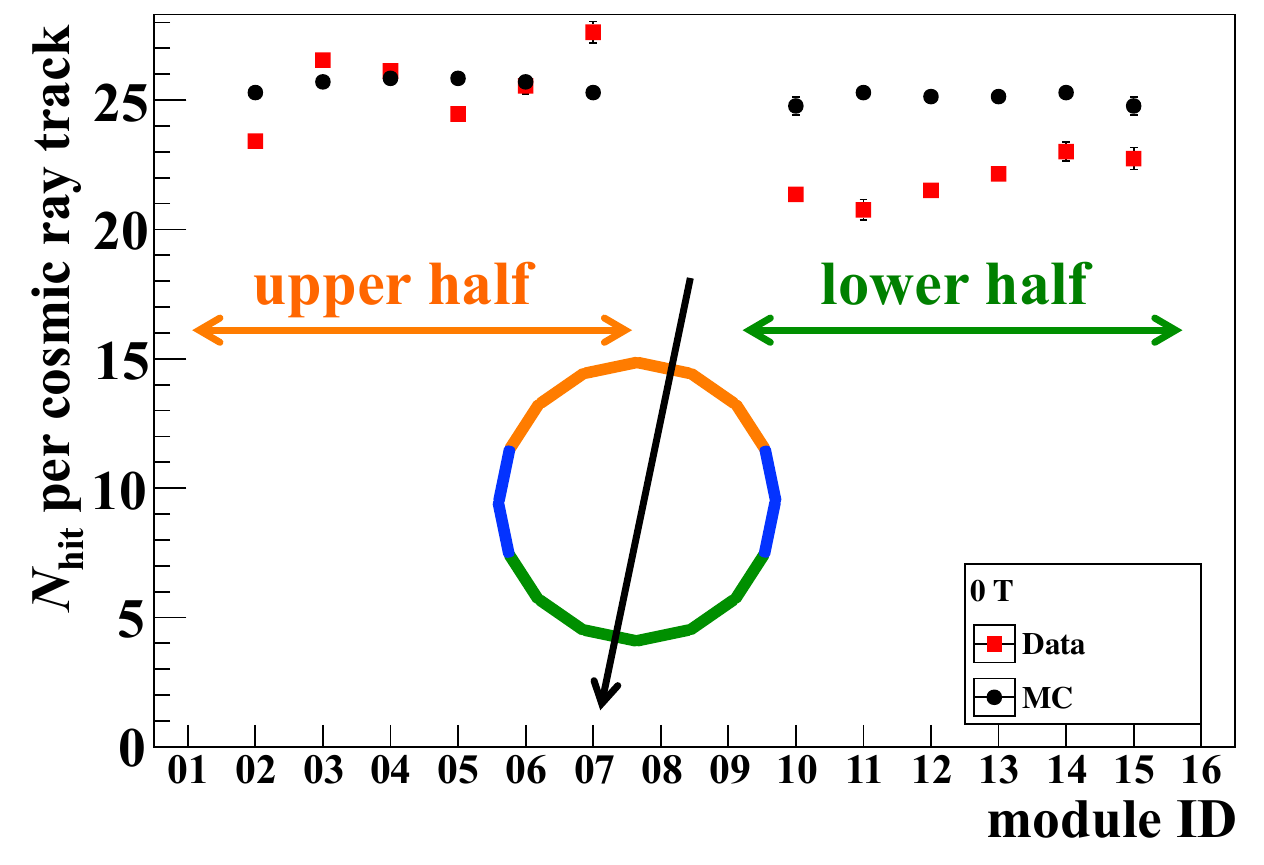}
\end{minipage}
\begin{minipage}{0.01\textwidth}
\end{minipage}
\begin{minipage}{0.49\textwidth}
\includegraphics[width=0.95\textwidth,bb=0 0 364 249,clip]{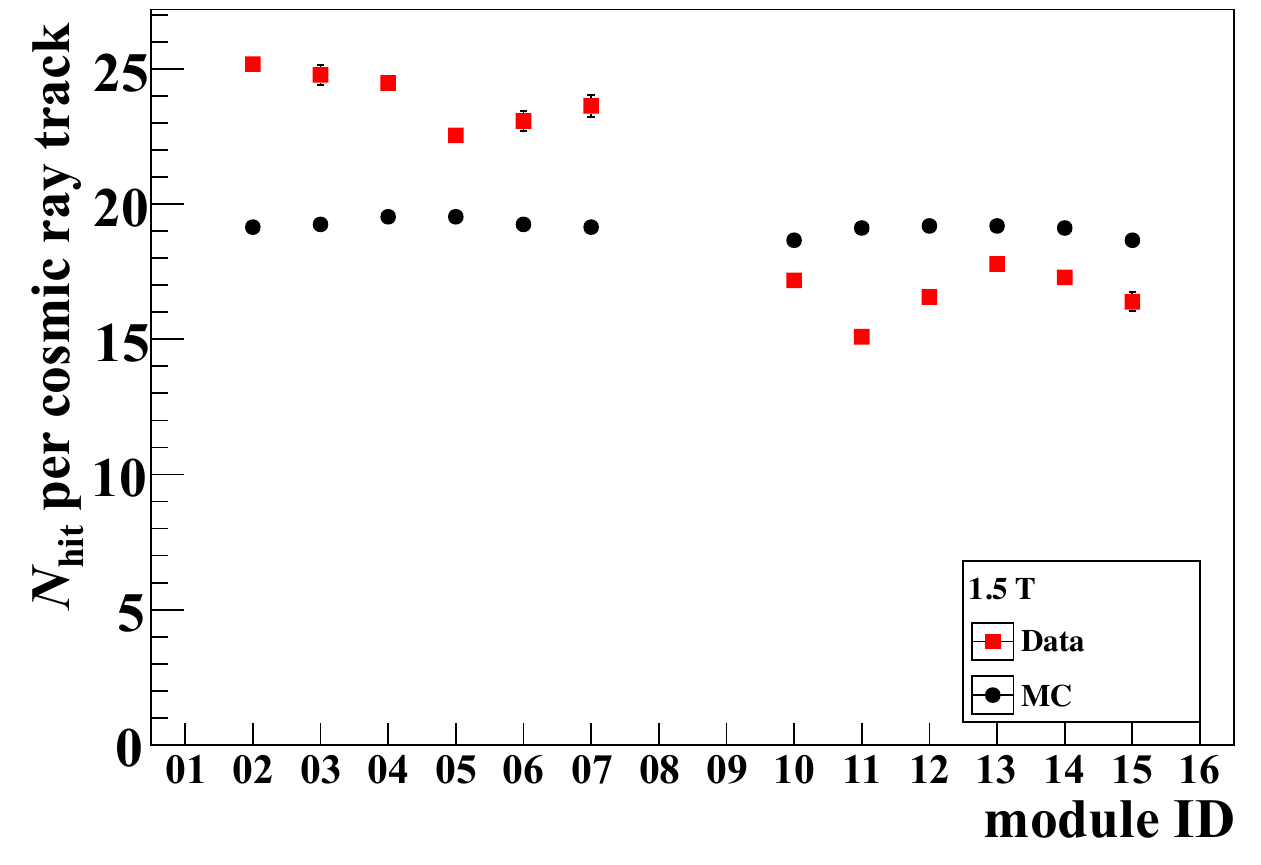}
\end{minipage}
\caption{\label{fig:CosmicNhit} Data-MC comparison of absolute $N_{\mathrm{hit}}$ values
		in the 0 T (left) and 1.5 T (right) magnetic field.
		Data for the side modules were not available because no trigger counters were set for these modules.}
\end{figure}

Data-MC comparison of the absolute $N_{\mathrm{hit}}$ values is shown in Fig.~\ref{fig:CosmicNhit},
where only diagonal pairs were considered.
The data $N_{\mathrm{hit}}$ was found to be consistent with MC simulation within 15\% in the 0-T condition.
For the 1.5-T case, discrepancy is at $20-30$\% level.
Still we did not see serious drop of $N_{\mathrm{hit}}$ in the magnetic field within this precision.
One of possible reasons of this discrepancy is
uncertainty of the angle and momentum distribution in the cosmic ray muon flux.
This is critical especially in the 1.5-T case
since impact angle strongly depends on these distribution2.
Another possibility is hit identification efficiency.
Gain characteristics are different in the upper and the lower half of the modules
as different types of PMTs \cite{MCPPMT} are used,
which could cause difference in hit efficiency of each Cherenkov photon.
Fine tuning of each PMT gain is in progress.

\section{Summary and prospects}

The Belle II Time Of Propagation counter is a novel device for particle identification,
where a Cherenkov ring image is reconstructed using timing information.
The detector was successfully installed into the Belle II structure
and commissioning is on-going.
Performance of the detector was evaluated by measuring the number of photon hits for cosmic ray muons,
which proved that the photon yield was consistent with simulation expectation
within 15(30)\% in operating without (with) the magnetic field.
More detailed performance studies are planned
by combining precise track information from the CDC detector \cite{CDC},
which was installed after the cosmic ray data taking. 


\acknowledgments

This work was supported by MEXT Grant-in-Aid for Scientific Research (S) ``Probing New Physics with Tau-Lepton.'' 


\end{document}